\documentclass[12pt]{article}
\usepackage{epsfig}

\newcommand{\tanbeta}    {$tan\beta$}
\newcommand{\beq}        {\begin{equation}}
\newcommand{\eeq}        {\end{equation}}
\newcommand{\beqa}        {\begin{eqnarray}}
\newcommand{\eeqa}        {\end{eqnarray}}

\begin{document}

\begin{flushright}
\large{hep-ph/9810307}
\end{flushright}
\vskip 1cm
\begin{center}
{\Large \bf
    Dileptonic decay of $B_s$ meson in SUSY models with large $tan\beta$
}

\vskip 3cm
{\bf
  S. Rai Choudhury\footnote{E-mail : src@ducos.ernet.in}
  and
  Naveen Gaur\footnote{E-mail : naveen@physics.du.ac.in} }
\\ 

{  Department of Physics \& Astrophysics\\
  University of Delhi \\
  Delhi - 110 007 \\
  India  \\ }
\end{center}

\vskip 1in
\begin{abstract}
	Decay of $B_s$ meson into the dileptonic channels of $e^+ e^-$
and $\mu^+ \mu^-$ are suppressed in the QCD corrected standard model
due to the vector and axial nature of the coupling to lepton
bilinears.We show that in SUSY theories with large $tan\beta$,
contributions to the amplitude arising out of exchange of neutral
Higgs Bosons considerably enhances the decay of $B_s$ into $l^+ l^-$. In
some region of parameter space, the enhancement is more than two
orders of magnitude, bringing it well within experimental
possibilities in the near future.
\end{abstract}

\pagebreak

	Flavor changing neutral current (FCNC) decays of the neutral
B-mesons , $B^0$ and $B_s^0$ provide unique testing grounds of the
Standard Model (SM) improved by QCD-corrections via the operator
product expansions{\cite{ali1}}. During the last few years, considerable
theoretical attention has therefore been focussed on decays like $B
\to K^* ~\gamma ,~~ B \to X_s ~\gamma,~~ B \to X_d ~l^+ ~l^-,
~~ B_s \to \gamma ~ \gamma$ and $ B_s \to \gamma l^+ ~l^-$ 
in view of the planned
experiments at B-factories, which are likely to measure branching
fractions as low as $10^{-8}${\cite{ali1}}. The calculations of the branching
ratios of FCNC B-decay processes are sensitive to new physics beyond
SM.In particular calculations based on Minimal Supersymmetric extension of
the Standard Model (MSSM) for many of
FCNC processes have been done{\cite{bert1},\cite{hewett}}.Nothing 
spectacularly different from SM
happens to the theoretical predictions but sizable changes in the
branching ratios can occur for some regions of the parameter space in
MSSM~{\cite{bert1},\cite{hewett}}.        

\indent The MSSM and in general any two Higgs-doublet model has in
addition to the unknown masses of the new particle content over and
above SM, the ratio of the vacuum expectation values of the two
neutral Higgs fields, ~$tan\beta = v_2/v_2$, as parameters ~{\cite{gunion}}.
From available data , on can only constrain the parameter space within
certain regions and typically $tan\beta$ can be constrained to be
$tan\beta \le 50$, {\it i.e.} it can be large ~{\cite{bert1},\cite{hewett}}.

\indent In some recent works{\cite{huang}}, it has been pointed out
that the processes $B \to X_s ~\tau^+ ~\tau^-$ and 
$B \to X_s ~\mu^+ ~\mu^-$
provide unique opportunities to distinguish SUSY models with large
 \tanbeta.~Exchange of Neutral Higgs Boson(NHB) gives
rise to new amplitude which were negligible in SM but which for large 
 \tanbeta~ can
enhance these processes by as much as $ 200 \%$. The purpose of this
note is to show that in SUSY-models with large \tanbeta ~ NHB exchange 
contributions causes a sizable enhancement of processes like $B_s \to
\mu^+ \mu^-$, as much as by two orders of magnitude in some allowed
region of parameter space.

\indent Let us start by recalling the result for $B_s \to \mu^+ \mu^-$
in QCD-improved Standard Model{\cite{ali2}}. The effective Hamiltonian
describing this process is :
\beqa
\label{one}
{\cal{H}}_{eff} 
&=& \frac{\alpha G_F}{\sqrt{2} \pi} \lambda
\Bigg[  
C^{eff}_9 ~({\bar s}~ \gamma^\mu~ P_L ~b)
({\bar l}~ \gamma^\mu ~l) \nonumber \\
& &
+ ~~C_{10}~({\bar s}~ \gamma^\mu~ P_L ~b)({\bar l} ~\gamma^\mu ~\gamma_5 ~l)
\nonumber \\
&&{}
+ \frac{2 C_7 m_b}{p^2} ({\bar s}~ \not\!p ~\gamma^\mu ~P_R ~b)
( {\bar l}~\gamma^\mu ~\gamma_5~ l) 
\Bigg]
\eeqa
with $\lambda \equiv V_{tb}~ V^*_{ts} ~, 
P_{L,R} = \frac{1}{2}~(1 \mp \gamma_5)$ and $p = p_+ + p_-$, the
sum of the momentum of $\mu^+$ and $\mu^-$. $C_7(m_b) ,~
C^{eff}_9(m_b)$ and $C_{10}(m_b)$ are the Wilson Co-efficients, whose
values are given by Misiak {\cite{misiak1}}.Since we are considering
$B_s$ , the matrix element of ${\cal H}_{eff}$ is to be taken between
vacuum and $|B^0_s\rangle$ state. This can be expressed in terms of the
$B^0_s$ decay constant $f_{B_s}~${\cite{lin}}.
\beqa
\label{two}
\langle 0~|~ {\bar s} ~\gamma^\mu~ \gamma_5 ~b~| B^0_s \rangle 
&=& - f_{B_s} p^\mu_B \\
\label{three}
\langle 0~| {\bar s} ~\gamma_5~ b| B^0_s \rangle 
&=& - f_{B_s} m_{B_s} 
\eeqa
and
\beq
\label{four}
\langle 0 |~ {\bar s} ~\sigma^{\mu \nu} ~P_R ~b~ |B^0_s\rangle = 0
\eeq
Since $ p^\mu_B = p^\mu_+ + p^\mu_-$, the $C_9$ term in eqn(\ref {one})
gives zero on contraction with the lepton bilinear, $C_7$ gives zero
by (\ref{four}) and the remaining $C_{10}$ term gets a factor of
$2~m_l$.
\par Thus 
\beqa
\label{four1}
\langle 0| {\cal H}_{eff} | B_s \rangle
&=& \frac{\alpha G_F}{\sqrt{2 \pi}} \lambda {1 \over 2} f_{B_s} C_{10}
p_B^\mu ~ \bar{l} \gamma_\mu \gamma_5 l    \nonumber \\
&=& \frac{\alpha G_F}{\sqrt{2 \pi}} \lambda f_{B_s} (C_{10} m_l)
~ (\bar{l} \gamma_5 l)
\eeqa

Decay rate of $B_s \to \mu^+ ~\mu^-$ is 
\beq
\label{five}
\Gamma(B_s \to l^+ ~l^-) 
= \frac{\alpha^2 G_F^2 f_{B_s}^2 m_{B_s} m_l^2}{16 \pi^3}
~|V_{tb}V^*_{ts}|^2 ~C_{10}^2
\eeq
The presence of the factor $m_l^2$ makes (\ref{five}) almost vanish
for $e^+ e^-$ and give rise to a small Branching ratio of $2.6\times
10^{-9}$ for $\mu^+ \mu^-$.

\indent Consider now the MSSM.~$l^+ ~l^-$ production via the neutral 
Higgs bosons
give rise to additional amplitudes and these become important for
large ~\tanbeta ~{\cite{huang}}.The neutral Higgs bosons coupling
to b-quark and $l^+$  proportional to $m_b tan\beta$ and
$m_l tan\beta$ respectively for large ~\tanbeta~{\cite{gunion}}.
As pointed out in {\cite{huang}} $m_l tan\beta$ can be
large $\sim m_b$ or more.~In addition, as has been pointed out in 
{\cite{huang}}, the $b \to s$ transition via chargino-stop loop gives
rise to another ~\tanbeta~ factor, so that the $b \to s ~l^+~ l^-$
amplitude goes like $m_b m_l (tan\beta)^3$ for large \tanbeta . 
Thus these additional amplitude
effectively do not suffer from the $m_l$ suppression of expression
(\ref{five}).The additional diagrams can be taken care of through and
additional effective Hamiltonian{\cite{huang}}
\beq
\label{six}
{\cal H}_{eff}^{NHB} = \frac{\alpha G_F}{\sqrt{2} \pi} ~\lambda
  ~ [~ C_{Q_1}~({\bar s} ~P_R ~b)({\bar l}~l) 
    + C_{Q_2}~({\bar s}~P_R~b)({\bar l} ~\gamma_5 ~l)~ ]
\eeq
The co-efficients $C_{Q_1}$,$C_{Q_2}$ are Wilson co-efficients.There
values at $M_W$ scale can be calculated by evaluating the Feynman
diagrams of NHBs contributions. The result is \cite{huang}
\beqa
\label{six1}
C_{Q_1}(m_W)
&=& \frac{m_b m_\mu}{4 m_{h^0}^2 sin^2\theta_W}
   tan^2\beta\{(sin^2\alpha + h cos^2\alpha)[ \frac{1}{x_{Wt}}
    (f_1(x_{Ht}) - f_1(x_{Wt}))   \nonumber \\
&+& \sqrt{2} \sum_{i=1}^{2} \frac{m_{\chi_i}}{m_W} \frac{U_{i2}}
     {cos\beta}( - V_{i1} f_1(x_{{\chi_1}\tilde{q}}) 
    + \sum_{k = 1}^{2} \Lambda(i,k)T_{k1} f_1(x_{\chi_1 \tilde{t}_k}))
    \nonumber \\
&+& (1 + \frac{m_{H_\pm}^2}{m_W^2}) f_2(x_{Ht},x_{Wt})] 
    - \frac{m_{h^0}^2}{m_W^2} f_2(x_{Ht},x_{Wt})    \nonumber  \\
&+& 2 \sum_{ii'}^2 ( B_1(i,i')\Gamma_1(i,i') +
A_1(i,i')\Gamma_2(i,i'))\} \\
\label{six2}
C_{Q_2}(m_W)
&=& - \frac{m_b m_\mu}{4 m_{A^0}^2 sin^2\theta_W}
   tan^2\beta\{[ \frac{1}{x_{Wt}}(f_1(x_{Ht}) - f_1(x_{Wt}))  
   + 2f_2(x_{Ht},x_{Wt})   \nonumber \\
&+& \sqrt{2} \sum_{i=1}^{2} \frac{m_{\chi_i}}{m_W} \frac{U_{i2}}
     {cos\beta}( - V_{i1} f_1(x_{{\chi_1}\tilde{q}}) 
    + \sum_{k = 1}^{2} \Lambda(i,k)T_{k1} f_1(x_{\chi_1 \tilde{t}_k}))
    \nonumber \\
&+& 2 \sum_{ii'}^2 (-U_{i'2}V_{i1}\Gamma_1(i,i') + U^*_{i2}V^*_{i'1}
\Gamma_2(i,i')) \}
\eeqa
where
\beqa
\label{six3}
B_1(i,i') &=&
( - {1 \over 2} U_{i'1} V_{i2} sin 2 \alpha (1 - h) 
  + U_{i'2}V_{i1}( sin^2\alpha + h cos^2 \alpha ))  \nonumber \\
A_1(i,i') &=&
 ( - {1 \over 2} U^*_{i1} V^*_{i'2} sin 2 \alpha (1 - h) +
U^*_{i2} V^*_{i'1}(sin^2\alpha + h cos^2\alpha))   \nonumber \\
\Gamma_1(i,i') &=& 
  m_{\chi_i}m_{\chi_{i'}} U_{i2}
   ( - \frac{1}{\tilde{m}^2} f_2(x_{\chi_i,\tilde{q}},x_{\chi_{i'}\tilde{q}})
    V_{i'1} 
  + \sum_{k=1}^2 \frac{1}{m_{\tilde{t}_k}^2} \Lambda(i',k)T_{k1}
f_2(x_{\chi_i\tilde{t}_k})) \nonumber \\
\Gamma_2(i,i') &=& 
   U_{i2}
   ( - f_2(x_{\chi_i,\tilde{q}},x_{\chi_{i'}\tilde{q}})
    V_{i'1} 
  + \sum_{k=1}^2 \Lambda(i',k)T_{k1}
f_2(x_{\chi_i\tilde{t}_k})) \nonumber \\
\Lambda(i,k) &=&
  V_{i1}T_{k1} - V_{i2} T_{k2} \frac{m_t}{\sqrt{2} m_W sin\beta}
\nonumber \\
f_1(x_{ij}) &=&
   1 -  \frac{x_{ij}}{x_{ij} - 1} ln x_{ij} + ln x_{Wj} \nonumber \\
f_2(x,y) &=&
  \frac{1}{x - y}( \frac{x}{x - 1} ln x - \frac{y}{y - 1} ln y )
\nonumber \\
x_{ij} &=& m_i^2/m_j^2 ~~~~~  ; ~~~~~ x_{Wj} = m_W^2/m_j^2
\eeqa
In (\ref{six1}),(\ref{six2}), the various m's represent the masses of
the corresponding particle and $ h = m_{h^0}^2/m_{H_0}^2$. U and V
represent the matrices which diagonalises the chargino masses.~T is the
matrix used for diagonalisation of stop mass matrix. $\tilde{m}$ is
the average mass of the first two generation of squarks. Starting with
$C_{Q_i}(m_W)$, there values at a scale $m_b$ can be evaluated by
solving the RGE equations.The result is 
\beq
\label{six4}
C_{Q_i}(m_b) = \eta^{- {\gamma_{Q_i}}/\beta_0} C_{Q_i}(M_W)
\eeq
where $\eta = \frac{\alpha_s(m_b)}{\alpha_s(M_W)} \approx
1.8$,$\gamma_{Q} = -4$ and $\beta = 11 - (2/3)n_f = 9$ \cite{cshuang}

 Taking the matrix elements of (\ref{six})
between vacuum and $|B^0_s\rangle$ as before, we see in effect that
$C_{Q_1}$ and $C_{Q_2}$ enter multiplied by $m_{B_s}$ unlike $C_{10}$ in
(\ref{five}) which entered as $m_l C_{10}$. For large ~\tanbeta~ thus,
the decay $B_s \to \mu^+ ~\mu^-$ will be dominated by (\ref{six}) and
we get :
\beqa
\label{seven}
\Gamma^{SUSY}(B_s \to \mu^+ ~\mu^-) 
&=&
\frac{\alpha^2 G_F^2 f_{B_s}^2 m_{B_s}}{16 \pi^3}~ |V_{tb}V^*_{ts}|^2
~\Bigg[~ \frac{C_{Q_1}^2 m_{B_s}^2}{4}           \nonumber \\
& & + (m_\mu C_{10} + \frac{m_{B_s}}{2}C_{Q_2})^2 ~\Bigg]
\eeqa
The co-efficient $C_{10}$ in (\ref{seven}) has been evaluated in NLO
approximation by Misiak \cite{misiak1} in SM.We use his value since SUSY
contributions are expected to change this value only slightly and also
because in the region of large $tan\beta$, the RHS of (\ref{seven})
will be dominated by the $C_{Q_1},C_{Q_2}$ term. As we can see from
(\ref{six1})-(\ref{six4}) the co-efficients
$C_{Q_1}$ and $C_{Q_2}$ depend, in addition to SM parameters,to the
MSSM parameters.The most economical version of MSSM is the one where
the spontaneous breaking of $SU(2)\bigotimes U(1)$ is achieved through
radiative effects.~Further at Unification scale of $M_{GUT}
(\sim 10^{16}$ GeV)
scalars have a common soft breaking mass term m and Gauginos also have
a unified soft breaking mass M.Thus at scale $M_{GUT}$,MSSM in
addition of SM parameters have five parameters : m,M, the ratio of the
vacuum expectation values of the two Higgs fields $tan\beta$,~the
trilinear soft breaking scalar term coefficient A and the bilinear
soft breaking term coefficient B.The values of these parameters at
scale $\sim m_b$ relevant to us can be related to their values at
$M_{GUT}$ through RGE which has been extensively discussed in
literature \cite{bert1},\cite{lahanas}.As pointed out in \cite{goto1}
to suppress the large contribution to $K^0 - \bar{K}^0$ mixing it is
sufficient to require degeneracy of the soft SUSY breaking mass in
squark sector.Thus, the strict universality of all scalar masses is
not necessarily required in context of SUGRA model.
We thus consider in our estimate of 
study $B_s \to \mu^+ \mu^-$ ,in regions of parameter space where 
 the condition of universality of scalar masses is relaxed.
\par The parameter space of MSSM is constrained by several pieces of
known experimental results, the $W^\pm,Z$ one loop mass corrections,
(g - 2) of the $\mu$ meson and most importantly by the lower and upper
bounds to $b \to s \gamma$ decay rate \cite{cleo} :
$${\cal B}(b \to s \gamma) = (0.6 - 5.4) \times 10^{-4}$$
MSSM parameter space in this context has been extensively analyzed by
Lopez et.al \cite{lopez},Goto et.al \cite{goto1}.
We present in Fig(\ref{relaxrate}) the rate of $\Gamma^{SUSY}(B_s
\to \mu^+ \mu^-)$ (\ref{seven}) relative to their SM values for some
of the allowed region of the parameter space for large values of
$tan\beta$ in SUGRA model where the universality of scalar masses in
SUGRA model is relaxed. The SM branching ratio of $B_s \to \mu^+
\mu^-$ is
$2.6 \times 10^{-9}$, so for large values of $tan\beta$ the branching 
ratio can be as large as $\sim 10^{-6}$.This is well within the 
experimental reach of
the B-factories expected soon.The $B_s \to \mu^+ \mu^-$ observation or
otherwise will thus serve as a very useful constraint on the parameter
space of MSSM.
\par 
We present the result of SUGRA model in which we are assuming the
universality of scalr masses in Fig(\ref{otherrate})
We also present the result of a more predictive
class of Supergravity (SUGRA) theories : the Moduli (or No-scale)
scenario ( m = A = 0 at unification scale) and the dilaton scenario
($m = M/\sqrt{3},A=-1$) in Fig(\ref{otherrate}).We see that with 
the allowed choice of
parameters the $B_s \to \mu^+ \mu^-$ rate depend crucially on the
value of the Higgs scalars and pseudo-scalars.For low values of those
masses with not too small splitting of the stops, the rate can be
enhanced by more than two orders of magnitude (Fig \ref{relaxrate}).
Thus $B_s \to
\mu^+ \mu^-$ data together with estimates of masses of superpartners
 will be very useful in cross-checking these models with
experimental results.
\par We next turn to the decay $B_s \to l^+ ~l^- ~\gamma$, which was
studied in the framework of QCD corrected effective SM Hamiltonian by
Eilam et.al.{\cite{eilam}}. As is pointed out there , the
radiative process unlike the non-radiative channel does not face the
helicity suppression factor proportional to $m_l$.The net result that
one ends up is paradoxical in that the Branching ratio for the
radiative mode turns out to be higher than the one for the non-radiative
mode notwithstanding the fact that the former has an extra factor of
$\alpha$. This is very easily seen by considering the set of graphs
where the photon line is hooked on to the various external lines of
the non-radiative decay process.Graphs when the photon line is hooked
on to the lepton gets a factor of $m_l$ and so can be ignored. The two
graphs with the photon hooked on to the b and s quark give rise to the
following amplitude for the process $B_s(p_{B_s}) \to \gamma(p_\gamma)
+ l^+(p_+) + l^-(p_-)$ {\cite{eilam}}
\beqa
\label{eight}
A(B_s \to l^+ ~l^- ~\gamma) 
&=& \frac{\alpha~ G_F}{\sqrt{2} \pi} \lambda 
 \epsilon_\nu ~
\Bigg[ \bar{s} ~[ 
 \gamma^\nu ~\frac{\not\!p_\gamma - \not\!p_s + m_s}{- 2 p_s.p_\gamma}
   ~\gamma^\mu ~P_L .~ Q_s         \nonumber \\
&+& P_R ~\gamma^\mu ~\frac{\not\!p_b - \not\!p_\gamma + m_b}{- 2 p_b.p_\gamma}
   ~ Q_b]~b~[C^{eff}_9 ~\bar{l}~\gamma_{\mu}~l 
          + C_{10}~\bar{l}~\gamma_\mu ~\gamma_5 ~l] \nonumber \\
&+& \frac{2 C_7 m_s}{p^2} ~[ P_R ~\not\!p ~\gamma^\mu
       \frac{\not\!p_b - \not\!p_\gamma + m_b}{- 2 p_b.p_\gamma} 
	~\gamma^\nu ~Q_b     \nonumber  \\
&+& \gamma^\nu ~\frac{\not\!p_\gamma - \not\!p_s + m_s}{- 2 p_s.p_\gamma}
    \not\!p ~\gamma^\mu ~P_R ~Q_s]~b~[\bar{l}~\gamma_\mu ~l]
\Bigg] 
\eeqa
In (\ref{eight}) $Q_s,~Q_b$ are the charges of b and s quarks , $p_b$
and $p_s$ respectively represent the momentum of the b and s quark
inside the $B^0_s$ and in the constituent quark model can be written
as $p_{b,s} = \frac{m_{b,s}}{m_{B_s}} ~p_{B_s}$.The fact that $m_s \ll m_b$
allows one to retain only the leading term in $(m_s/m_b)$ in
eqn(\ref{eight}) and gives to the amplitude:
\beqa
\label{nine}
A(B_s \to l^+ ~l^- ~\gamma) 
&\simeq&
  \frac{e ~\alpha G_F f_{B_s} \lambda}
       {12 \sqrt{2} \pi [~\frac{m_s}{m_b}(p_B.p_\gamma)]}
  [ p^\nu_B \{ (p_\gamma)^\mu \epsilon_\nu 
             - (p_\gamma)_\nu \epsilon^\mu \}          \nonumber \\
& &   + ~\it{i} \epsilon^{\nu \alpha \mu \beta}
         (p_\gamma)_\alpha (p_B)_\beta ~]               \nonumber   \\
& &
  [~ (C_9 - \frac{2 C_7 m_{B_s}^2}{p^2})(\bar{l}~\gamma_\mu ~l)
   + C_{10}(\bar{l}~\gamma_\mu ~\gamma_5 ~l)~]
\eeqa 
Where once again we have used
eqns(\ref{two}),(\ref{three}),(\ref{four}).Using the values of
$C_9,C_7$ and $C_{10}$ the Branching ratio into the $B_s^0 \to \mu^+
\mu^- \gamma$ mode is estimated at $4.6 \times 10^{-9}$ {\cite{eilam}}
, almost twice
the branching ratio of $B_s^0 \to \mu^+ \mu^-$. \\
\indent Let us now examine the effect of the NHB exchange diagram in
SUSY models on the amplitude $A(B_s \to l^+ l^- \gamma)$.The
contribution can be written in terms of $C_{Q_1}$ and $C_{Q_2}$ terms.
So $A(B_s \to l^+ ~l^-~\gamma)$ is :
\beqa
\label{ten}
A^{NHB}(B_s \to l^+ l^- \gamma) 
&\simeq&
\frac{\alpha G_F}{12 \sqrt{2} \pi} f_{B_s} \lambda \epsilon_\nu 
  \bar{s}~[ \gamma^\nu ~\frac{\not\!p_\gamma}
{\frac{m_s}{m_b}p_B.p_\gamma}.P_R~]~b
\nonumber \\
& &  [C_{Q_1} (\bar{l}l) + C_{Q_2}(\bar{l}\gamma_5 l)]
\eeqa
where we have made the same approximations as in deriving
eqn(\ref{nine}) {\it i.e.} retaining only the leading power of $(m_s/m_b)$.\\ 
However :
\beqa
\label{ten1}
\epsilon_\nu \langle 0 | \bar{s} \gamma^\nu \not\!p_\gamma P_R b
| B_s^0\rangle 
&=& {1 \over 2} \epsilon_\nu p^\nu ~ \langle 0|\bar{s} P_R b | B_s^0
\rangle         \nonumber       \\
&& + {1 \over 2} ~\epsilon_\nu p_\gamma^\mu ~ \langle 0 | \bar{s}
[\gamma^\nu,\gamma^\mu]P_R b| B_s^0 \rangle  \nonumber \\
&=&  0
\eeqa 
from eqns.(\ref{two}) and the transversality condition for real photon
$\epsilon_\nu p_\gamma^\nu = 0$.

Thus,the NHB terms are
negligible even for large \tanbeta ~co-efficients $C_9$ and $C_{10}$ do
not change much in MSSM.~Changes in $C_7$ in SUSY are constrained by $b
\to s \gamma$ data.This change too is less than a factor of two.Thus
the SM estimate of (\ref{nine}) effectively still holds in MSSM also.
\par We thus can have a change in the pattern of conclusion of
\cite{eilam} when the Neutral Higgs Boson contribution are included
with the large \tanbeta depending upon SUSY and most importantly upon
the values of Higgs masses and \tanbeta 
.The radiative mode $B_s^0 \to l^+ l^- \gamma$
branching ratio stay put at about $4.6 \times 10^{-9}$, whereas the
non-radiative(suppressed earlier without NHB) can jump by a factor of the
order $\sim 10^2$ to a value $\sim 10^{-7}$.With the flux of $B_s^0$
expected to go up to observe such branching ratios, one thus has a concrete
and interesting situation to test SUSY-models with large \tanbeta ~ in
studying the radiative and non-radiative dileptonic decays of $B^0_s$


\bibliographystyle{plain}

\pagebreak



\begin{center}
{\bf FIGURE CAPTIONS}
\end{center}
{\bf Figure 1} :
Wilson coefficient $C_{Q_1}$ as a function of light chargino mass in
SUGRA model without nonuniversality of scalar masses (pseudo-scalar higgs
mass is taken to be in range $80 < m_A < 700$),
m = M, A = - 1/2.     \\
{\bf Figure 2} :
Wilson coefficient $C_{Q_2}$ as a function of light chargino masses in
SUGRA model without nonuniversality of scalar masses with same
parameters as above.Parameters are $80 < m_A < 700, m = M , A = - 1/2$  \\
{\bf Figure 3} : 
Ratio of $\Gamma^{SUSY}/\Gamma^{SM}$ (SUSY rate normalized to SM rate)
as a function of light chargino mass in SUGRA model without
nonuniversality of scalar masses.Parameters are $80 < m_A < 700, m = M
, A = - 1/2 $                          \\
{\bf Figure 4} :
$\Gamma^{SUSY}/\Gamma^{SM}$ as a function of light chargino mass in \\
(a) Dilaton (m = M/$\sqrt{3}$ , A = - 1) \\
(b) SUGRA ( m = M , A = - 1/2 ) \\
(c) Moduli ( m = A = 0) \\
Points with dots (.) are consistent with allowed parameter space
whereas points marked with plusses (+) are those excluded by 
our choice of parameter space (including $b \to s \gamma$ constraints).
\pagebreak

\begin{figure}
\epsfig{file=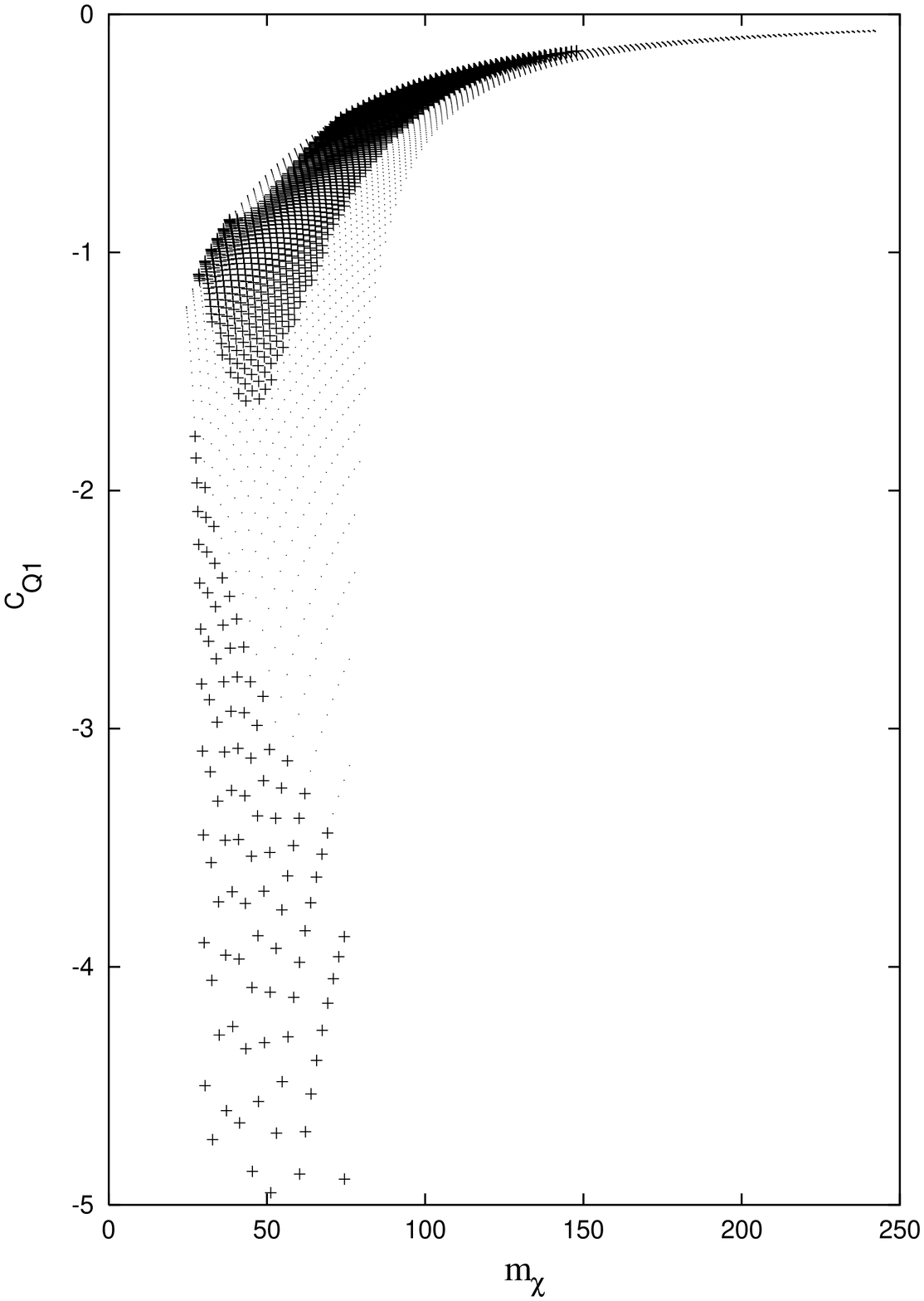,width=15cm,height=18cm}
\caption{}
\label{relaxcq1}
\end{figure}
\begin{figure}
\epsfig{file=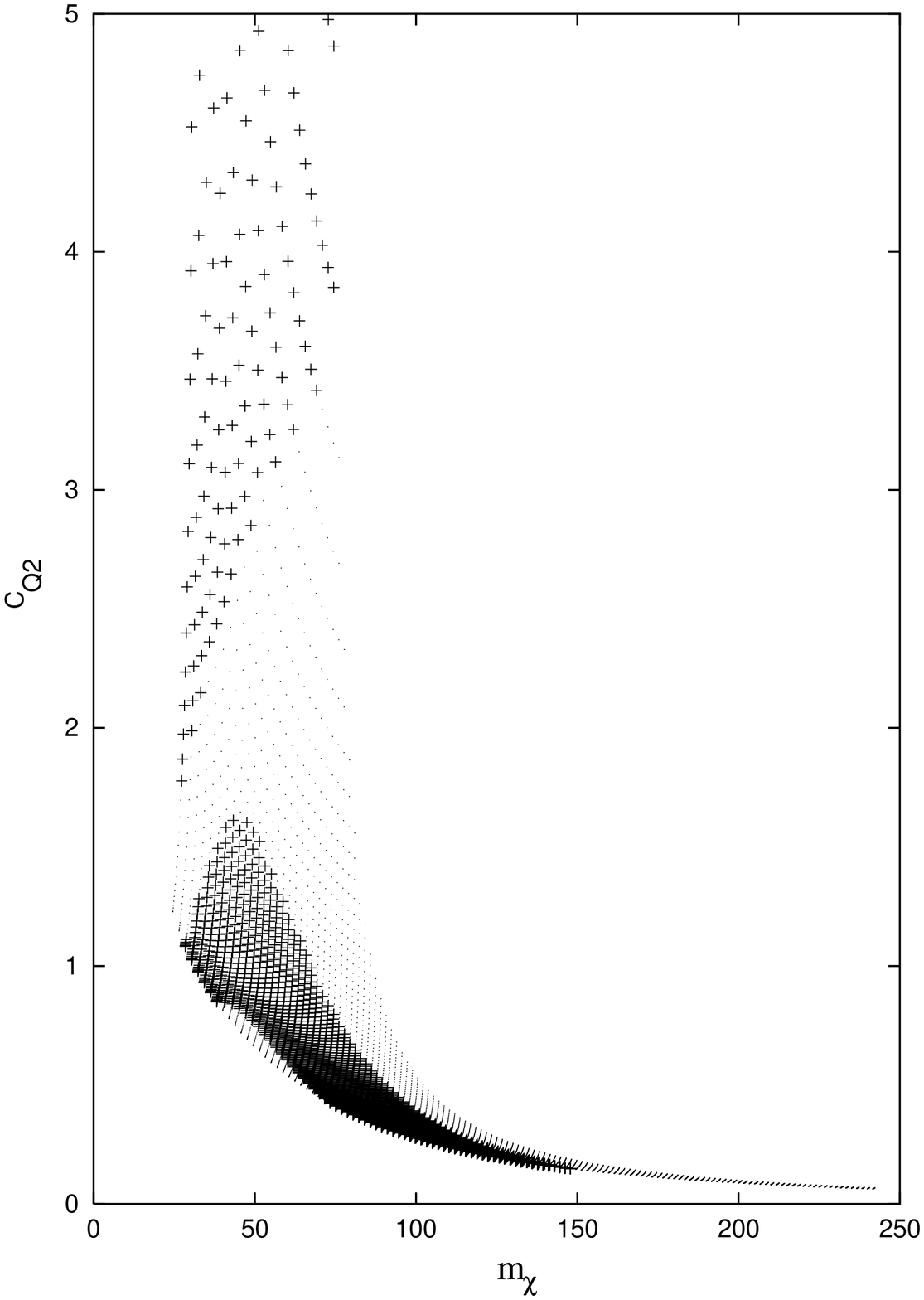,width=15cm,height=18cm}
\caption{}
\label{relaxcq2}
\end{figure}
\begin{figure}
\epsfig{file=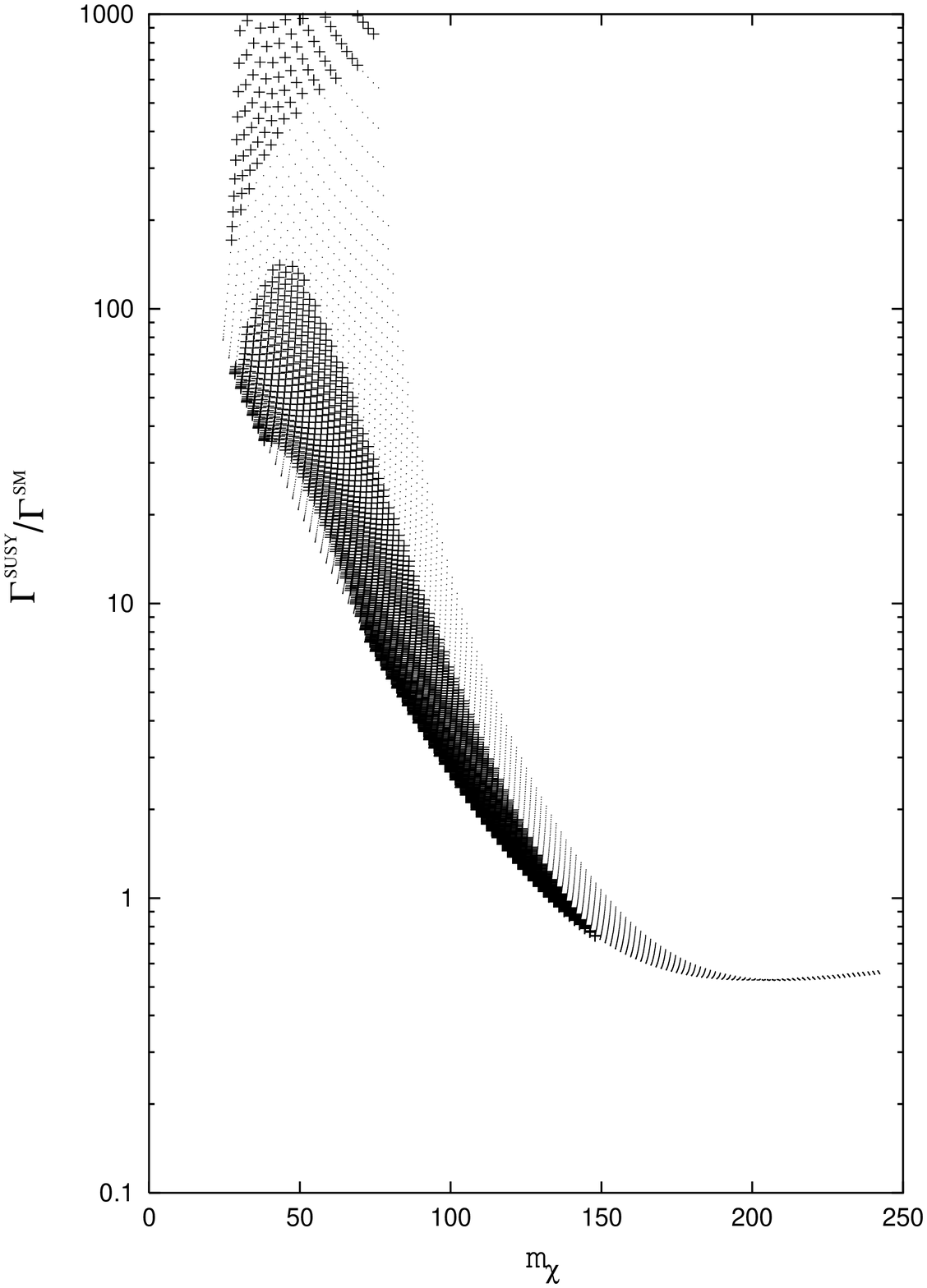,width=15cm,height=18cm}
\caption{}
\label{relaxrate}
\end{figure}
\begin{figure}
\epsfig{file=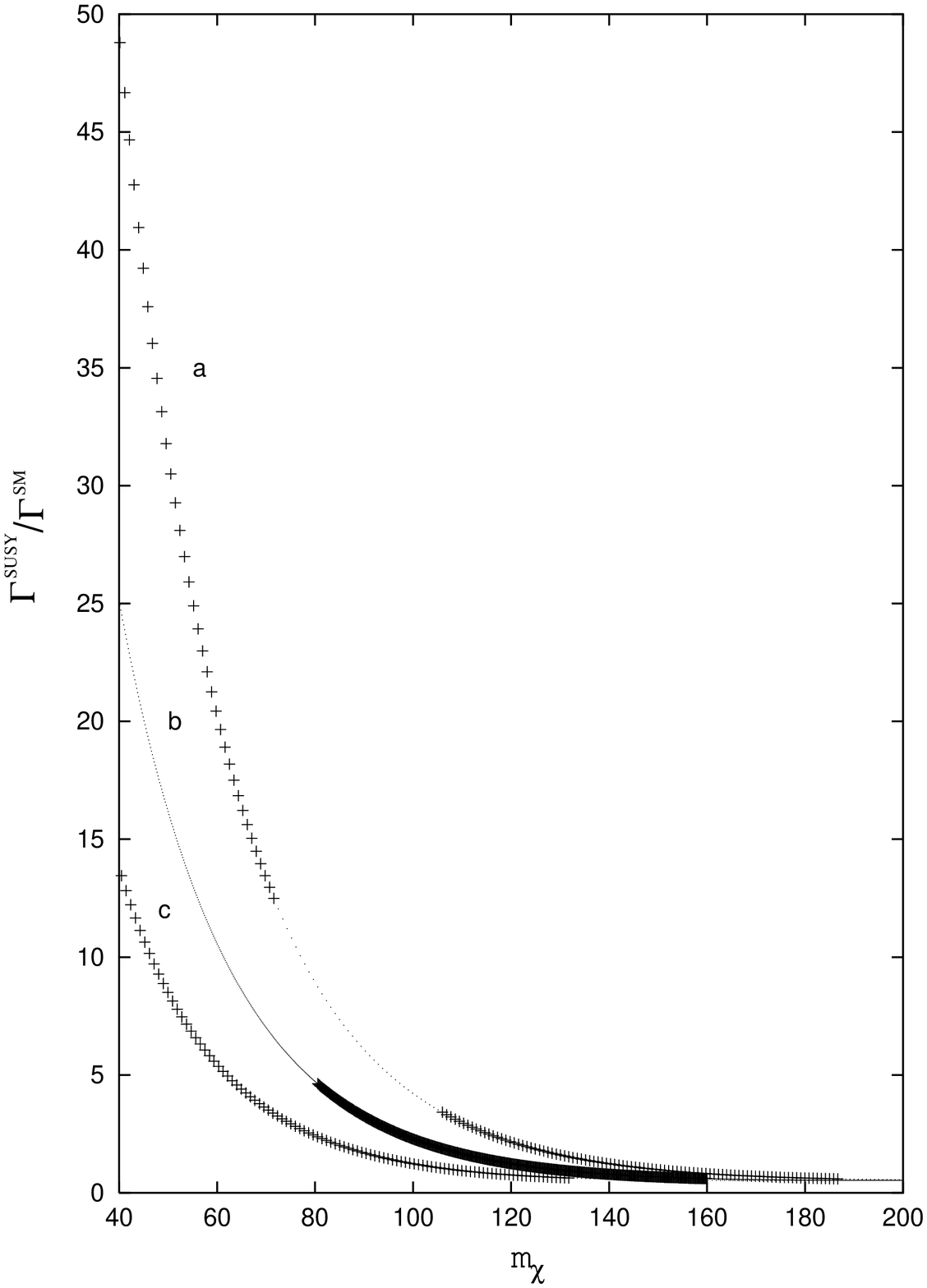,width=15cm,height=18cm}
\caption{}
\label{otherrate}
\end{figure} 

\end{document}